\documentclass{sf2a-conf}
\usepackage{graphicx}

\begin{document}
\TitreGlobal{SF2A 2007}
\title{Analytical Solutions of specific classes of Astrophysical Radiating Shocks}

\author{E. Falize$^{1,}$} \address{D\'epartement de Physique Th\'eorique et Appliqu\'ee, CEA/DIF, BP 12 91680 Bruy\`eres-le-Ch\^atel, France}, \address{LUTH, Observatoire de Paris, CNRS, Universit\'e Paris Diderot ; 5 Place Jules Janssen, 92190 Meudon, France} 
\author{C. Michaut$^2$}
\author{S. Bouquet$^{1, 2}$} 
\author{C. Cavet$^2$}

\runningtitle{Analytical Solutions of Astrophysical Radiating Shocks}

\setcounter{page}{1}

\index{Falize E.}
\index{Bouquet S.}
\index{Michaut C.}
\index{Cavet C.}
\maketitle

\begin{abstract}
In this paper we study specific classes of radiating shocks which are widely spread in astrophysical environments. We present more general solutions of their structure and proceed to the analytical determination of physical quantities.
\end{abstract}
%
\section{Introduction}
Radiating shocks play a crucial role in astrophysical environments (Mignone 2005)  as well as in laboratory plasmas generated by powerful facilities (Drake 2006). In astrophysics, we can find them in the head of stellar and galactic jets (Blondin \& Cioffi 1989), in the first phase of supernovae explosions, in the late phase of supernova remnants (Chevalier \& Blondin 1995), in accreting systems such as the magnetospherical accretion of T Tauri (Gunther et al. 2007) or magnetic cataclysmic variables (Cropper 1990). These shocks are at the basis of several models which allow the interpretation of astronomical observations. For instance, they may explain the recent observations of X spectra (Gunther et al. 2007) of classical T Tauri: TW Hya, BP Tau and V4056Sgr. Thus, it is very important to understand the structure and stability of these shocks. The structure of radiating shocks depends on the nature of pre-shock and post-shock media (Drake 2006). In this paper, we will consider shocks for which radiative effects can be modelled by entropy losses. This is the more common shock that it is encountered in interstellar phenomena. We will derive a solution by assuming a power law model for the cooling function, which generalizes the five specific solutions we find in the literature (Chevalier \& Imamura 1982). Firstly, we will recall the theoretical model and discuss the approximation. Secondly, we will present the analytical solution for the one- and two-cooling processes problems. Finally, we proceed to the evaluation of fundamental physical quantities that characterize this type of shock and compare them with the results we already have.   
\section{Approximations and theoretical post-shock medium modelling}
We consider a plane-parallel $(\partial / \partial y=\partial / \partial z=0)$ collisional shock (\emph{i.e.}, $t_{ii}<< t_{dyn}$ where $t_{ii}$ is the characteristic time of collision between ions and $t_{dyn}$ is the dynamical time) with a post-shock medium which can be defined by a single temperature model (\emph{i.e.}, $t_{ei}<< t_{cool}$ where $t_{ei}$ is the characteristic time of energy exchanged between electrons and ions and $t_{cool}$ is the radiative cooling time). Moreover, we suppose that the local gravitational field does not modify the shock structure (\emph{i.e.}, $g_{*}x_{s}<<v_{s}^{2}$ where $g_{*}$, $x_{s}$ and $v_{s}$ are respectively the gravitational field, the thickness of the cooling layer and the shock velocity). Finally, we assume that the shock is stationnary (\emph{i.e.}, $t_{cool}<<t_{dyn}$ which means that cooling effects are faster than dynamical ones). Thus, the equations that give the evolution of post-shock medium are (Kylafis \& Lamb 1982):
\begin{equation}\label{eq1:eq}
\frac{d}{dx}[\rho v]=0 \quad \frac{d}{dx}[\rho v^{2}+P]=0 \quad v\left[\frac{dP}{dx}-\gamma \frac{P}{\rho}\frac{d\rho}{dx}\right]=-(\gamma-1)\Lambda(\rho,P,x)
\end{equation}
where $x$, $\rho$, $v$, $P$, $\gamma$, $\Lambda$ are respectively the spatial coordinate, the density, the velocity, the pressure, the polytropic index and the cooling function. Although these equations are theoretically consistent, we must specify a equation of state in order to connect the microscopic phenomena to the cooling function. We consider: $P=\epsilon[Z]\rho^{\alpha}T^{\beta}$ where T, $\epsilon[Z]$, $\alpha$ and $\beta$ are respectively the temperature, a function of ionization Z and two free exponants where we must impose  $\gamma(1-\beta)=(\alpha-\beta)$ in order to make sure that the entropy is preserved in the post-shock region.
\begin{figure}[h]
\begin{center}
\resizebox{9cm}{!}{\includegraphics{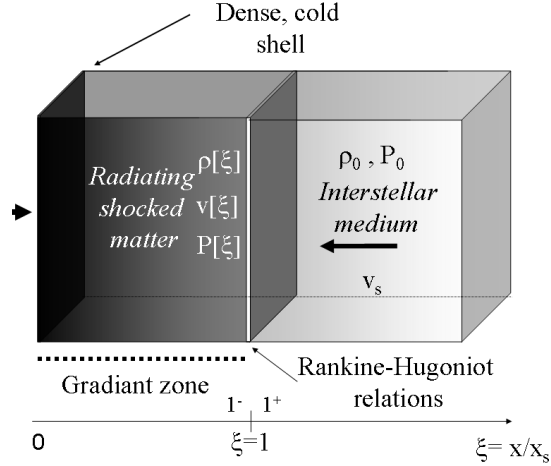}} 
\end{center}
\caption{Schematic representation of radiating shock with the definition of the different physical quantities and notations.} 
\label{myfig1} 
\end{figure} 
 On the shock front we have the Rankine-Hugoniot relations which are satisfied:
\begin{equation}
\frac{\rho[x_{s}^{-}]}{\rho_{0}}=\frac{(\gamma+1)M^{2}}{(\gamma-1)M^{2}+2},\quad \left|\frac{v[x_{s}^{-}]}{v_{s}} \right|=\frac{(\gamma-1)M^{2}+2}{(\gamma+1)M^{2}}, \quad \frac{P[x_{s}^{-}]}{\rho_{0}v_{s}^{2}}=\frac{2\gamma M^{2}-(\gamma-1)}{\gamma(\gamma+1)M^{2}}
\end{equation}
where $\rho_{0}$ and $M$ are respectively the pre-shock density and the Mach number (see the figure 1). It appears that a natural rescaling is the normalisation of spatial coordinate to the thickness of the cooling zone ($\xi=x/x_{s}$), the density in the post-shock region to the pre-shock one, the velocity to the shock velocity and the pressure to the ram pressure. In order to reduce the number of equations (\ref{eq1:eq}) we introduce an intermediate function  $\eta[\xi]$ defined by $v[\xi]=-v_{s}\eta[\xi]$. The first and second equations of (\ref{eq1:eq}) imply $\rho[\xi]=\rho_{0}/\eta[\xi]$ and $P[\xi]=\rho_{0}v_{s}^{2}(a_{1}\eta[\xi]+a_{2})$ where $a_{1}$ and $a_{2}$ are two constants. This transformation is consistent with the Rankine-Hugoniot relations provided $\eta[1]=[2+(\gamma-1)M^{2}]/[(\gamma+1)M^{2}]$ and $P[\xi]=\rho_{0}v_{s}^{2}(1+1/\gamma M^{2}-\eta[\xi])$. The determination of the intermediate function is possible through the last equation of (\ref{eq1:eq}) which is given by:       
\begin{equation}\label{eq3:eq}
\left[(\gamma+1)\eta-\gamma \left(1+\frac{1}{\gamma M^{2}}\right)\right]\frac{d\eta}{d\xi}=-(\gamma-1)\frac{x_{s}}{\rho_{0}v_{s}^{3}}\Lambda(\rho,P,\xi)
\end{equation}
It is this equation that we examine now.

\section{Analytical solutions to one- and two-power laws cooling functions problem}
First, we suppose that the cooling function writes: $\Lambda[\rho,P,x]=\Lambda_{0}\rho^{\epsilon}P^{\zeta}(x+X_{0})^{\theta}$ where $\Lambda_{0}$, $\epsilon$, $\zeta$, $\theta$ and $X_{0}$ are five constants. This form generalizes the optically thin case where $\theta=0$. The arbitrarity of $\epsilon$ and $\zeta$ is motivated by the fact that $\Lambda\propto \kappa_{P}\sigma T^{4}$ where $\kappa_{P}$ is the Planck opacity which can be modelled by a power law at high temperature. Introducing this form in equation (\ref{eq3:eq}) and using the integral form of Gauss hypergeometric function (that we note $F_{21}[a,b;c;x]$), we obtain an implicit form of the compression profile:   
\begin{equation}
cste+(\gamma+1)\eta^{\epsilon+2}\left[1+\frac{1}{\gamma M^{2}}\right]^{-\zeta}\frac{\Gamma[\epsilon+2]}{\Gamma[\epsilon+3]}F_{21}\left[\zeta,\epsilon+2;\epsilon+3;\frac{\eta}{1+1/[\gamma M^{2}]}\right]-
\end{equation}
\begin{equation}
\gamma\eta^{\epsilon+1}\left[1+\frac{1}{\gamma M^{2}}\right]^{1-\zeta}\frac{\Gamma[\epsilon+1]}{\Gamma[\epsilon+2]}F_{21}\left[\zeta,\epsilon+1;\epsilon+2;\frac{\eta}{1+1/[\gamma M^{2}]}\right]=\left\{ \begin{array}{ll}
\kappa_{0}ln[\xi+X_{0}/x_{s}]; \quad \theta = -1\\
\frac{\kappa_{0}}{\theta+1}\left[\xi+X_{0}/x_{s}\right]^{\theta+1}; \quad \theta \neq -1  \end{array}\right.
\end{equation}
where $\kappa_{0}=(\gamma-1)x_{s}^{\theta+1}\Lambda_{0}\rho_{0}^{\epsilon+\zeta-1}v_{s}^{2\zeta-3}$ and $cste$ is defined by a specific value of the implicit function. This general solution is an extension of the five specific solutions known in the literature (Chevalier \& Imamura 1982). 
Now we consider the case with two optically thin cooling processes. The cooling function writes: $\Lambda(\rho,P)=\Lambda_{0,1}\rho^{\epsilon_{1}}P^{\zeta_{1}}+\Lambda_{0,2}\rho^{\epsilon_{2}}P^{\zeta_{2}}$ where we suppose, without loss in generality, that $t_{cool,1}<t_{cool,2}$. Thus, the development of inverse cooling function gives: 
\begin{equation}
\frac{1}{\Lambda(\rho,P)}=\frac{1}{\Lambda_{0,1}}\left[\sum_{n=0}^{\infty}(-1)^{n}\left(\frac{\Lambda_{0,2}}{\Lambda_{0,1}}\right)^{n}\rho^{[\epsilon_{2}-\epsilon_{1}]n-\epsilon_{1}}P^{[\zeta_{2}-\zeta_{1}]n-\zeta_{1}}\right]
\end{equation}
which corresponds to a superposition of an infinity of processes. Introducing this expression into (\ref{eq3:eq}) and using once more the integral representation of Gauss hypergeometric function leads to an implicit solution: 
$$
cste+\sum_{n=0}^{\infty}[-1]^{n}\Lambda_{0,n}\left\{(\gamma+1)\eta^{\epsilon_{n}+2}\left[1+\frac{1}{\gamma M^{2}}\right]^{-\zeta_{n}}\frac{\Gamma[\epsilon_{n}+2]}{\Gamma[\epsilon_{n}+3]}F_{21}\left[\zeta_{n},\epsilon_{n}+2;\epsilon_{n}+3;\frac{\eta}{1+1/[\gamma M^{2}]}\right]-\right.
$$
\begin{equation}
 \left.\gamma\eta^{\epsilon_{n}+1}\left[1+\frac{1}{\gamma M^{2}}\right]^{1-\zeta_{n}}\frac{\Gamma[\epsilon_{n}+1]}{\Gamma[\epsilon_{n}+2]}F_{21}\left[\zeta_{n},\epsilon_{n}+1;\epsilon_{n}+2;\frac{\eta}{1+1/[\gamma M^{2}]}\right]\right\}= \frac{x_{s}}{\rho_{0}v_{s}^{3}} \xi
\end{equation}
where $\epsilon_{n}=\epsilon_{1}-[\epsilon_{2}-\epsilon_{1}]n$, $\zeta_{n}=\zeta_{1}-[\zeta_{2}-\zeta_{1}]n$ and $\Lambda_{0,n}=([\Lambda_{0,2}/\Lambda_{0,1}]^{n}/\Lambda_{0,1})\rho_{0}^{-\epsilon_{n}}[\rho_{0}v_{s}^{2}]^{-\zeta_{n}}$
We can see this solution as the one-process solution with a correction due to the presence of a second process.

\section{Analytical evaluation of fundamental physical quantities}

\begin{figure}[h]
\begin{center}
\resizebox{9cm}{!}{\includegraphics{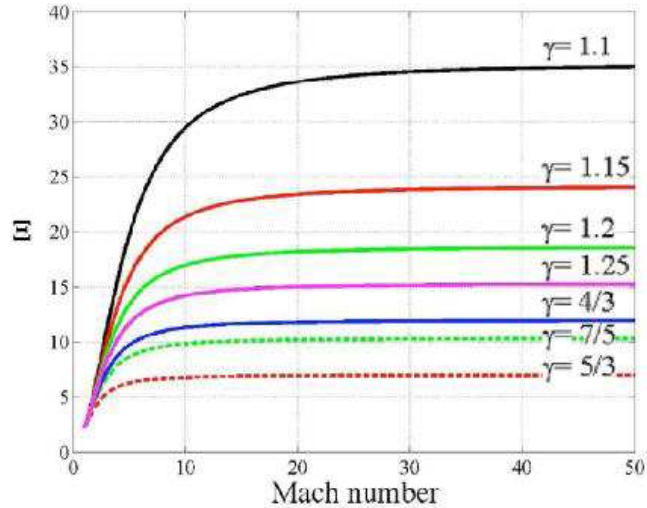}} 
\end{center}
\caption{Representation of the accreted column density versus Mach number for bremsstrahlung cooling for several values of polytropic index. We observe the saturation which is comptatible with values of table 1.} 
\label{myfig2} 
\end{figure}

We will evaluate two fundamental quantities which are the thickness (which corresponds to $x_{s}$) of the post-shock medium and the accreted column density (that we note $\Xi$). From the previous results, it is easy to show that:
$$
x_{s}=\frac{\rho_{0}v_{s}^{3}}{\Lambda_{0}(\gamma-1)\rho_{0}^{\epsilon}[\rho_{0}v_{s}^{2}]^{\zeta}}\left\{\gamma\left(\frac{2+(\gamma-1)M^{2}}{(\gamma+1)M^{2}}\right)^{\epsilon+1}\left[1+\frac{1}{\gamma M^{2}}\right]^{1-\zeta}\frac{\Gamma[\epsilon+1]}{\Gamma[\epsilon+2]}\times\right.
$$
\begin{equation}\label{eq4:eq}
\left. F_{21}\left[\zeta,\epsilon+1;\epsilon+2;\frac{2+(\gamma-1)M^{2}}{(\gamma+1)M^{2}(1+1/[\gamma M^{2}])}\right]-(\gamma+1)\left(\frac{2+(\gamma-1)M^{2}}{(\gamma+1)M^{2}}\right)^{\epsilon+2}\left[1+\frac{1}{\gamma M^{2}}\right]^{-\zeta}\frac{\Gamma[\epsilon+2]}{\Gamma[\epsilon+3]}\times\right.
\end{equation}
$$\left. F_{21}\left[\zeta,\epsilon+2;\epsilon+3;\frac{2+(\gamma-1)M^{2}}{(\gamma+1)M^{2}(1+1/[\gamma M^{2}])}\right] \right\} = \Delta[\gamma,M,\epsilon,\zeta] \times v_{s}t_{cool}$$
We can apply this result to magnetic cataclysmic variables in which $v_{s}=\sqrt{2GM_{wd}/R_{wd}}$ where $G$, $M_{wd}$ and $R_{wd}$ are respectively the gravitational constant, the white dwarf mass and the white dwarf radius. Furthermore, the accretion rate $\dot{m}$ is given by $\dot{m}=\rho_{0}v_{s}$. The introduction of these relations in equation (\ref{eq4:eq}) with assumptions of strong shock ($M\rightarrow \infty$) and $\gamma=5/3$ provides a simple relation between $x_{s}$, $\dot{m}$, $M_{wd}$ and $R_{wd}$:
\begin{equation}
x_{s}=7.60\times 10^{6}cm \left[\frac{\dot{m}}{4 g.cm^{-2}.s^{-1}}\right]^{-1}\left[\frac{M_{wd}}{0.5 M_{\odot}}\right]^{3/2}\left[\frac{R_{wd}}{10^{9}cm}\right]^{-3/2}
\end{equation}
This relation is consistent with Wu et al. (1994) results.
\begin{table}[h]
\caption{Satured values (limit $M\to \infty$) of the $\Xi$-parameter for different cooling processes} 
\label{table:1} 
\centering 
\begin{tabular}{c c c c c c c c c c} 
\hline\hline 
$(\epsilon,\zeta)$ \quad / $\gamma$ & $5/3$ & $7/5$ & $4/3$ & $1.1$ & $1.06$  \\ 
\hline 
( -2.35, 2.50) & 1.991 & 2.894  & 3.406 & 11.387 & 19.022 \\ 
( 1.50, 0.50) &  6.967 &  10.296 & 11.961 & 35.288 & 57.509\\ 
( 2.50, -0.50) &  5.869& 8.651 & 10.047 & 29.629 & 48.292 \\ 
( 2.00, 0.00) &  6.286 & 9.273 & 10.769 & 31.756 & 51.752 \\ 
( 1.00, 1.00) &  8.302 &  12.313 & 14.316 & 42.327 & 68.994\\ 
( 3.00, -1.00)&  5.585 & 8.232& 9.560 & 28.208 & 45.983 \\
\hline 
\end{tabular} 
\end{table} 
Now we can evaluate the accreted column. By definition, we have $\Xi=\left(\int_{0}^{x_{s}}\rho(x).dx\right)/\left(\int_{0}^{x_{s}}\rho_{0}.dx\right)$. This estimation is achieved for the former five analytical solutions by Laming (2004). Our results are presented in  figure \ref{myfig2} and numerical results  in table \ref{table:1}. These results generalize Laming ones for any cooling function. We find the trivial result which is that, for a given radiating process, smaller the polytropic index is, larger accreted matter is.

\section{Conclusion}
In this paper we present the general analytical solution to the problem of astrophysical radiating shock in the case where we have one- or two-cooling processes. These results are very important to validate
the input density profile which is at the basis of all stability studies. Our solutions are totally consistent with previous evaluation of several physical quantities. Moreover it makes possible to predict more general results. 
This study can be considered as the counterpart of studies relative to optically thick radiative shocks (Bouquet et al. 2000). 


\begin{thebibliography}{}

\bibitem{}Blondin, J. M., \& Cioffi, D. F. 1989, ApJ, 345, 853
\bibitem{}Bouquet, S., Teyssier, R., \& Chieze, J.-P. 2000, ApJS, 127, 245
\bibitem{}Chevalier, R. A., \& Imamura, J. N. 1982, ApJ, 261, 543
\bibitem{}Chevalier, R. A., \& Blondin, J. M.  1995, ApJ, 444, 312
\bibitem{}Cropper, M. 1990, Space Science Reviews, 54, 195
\bibitem{}Drake, R. P.  2006, {\it High-Energy-Density Physics} {\bf Springer-Verlag}
\bibitem{}Gunther, H. M., et al.  2007, A\&A, 466, 1111
\bibitem{}Kylafis, N. D., \& Lamb, D. Q. 1982, ApJ Suppl. Series, 48, 239
\bibitem{}Laming, J. M. 2004, Phys. Rev. E, 70, 057402
\bibitem{}Mignone, A. 2005, ApJ, 626, 373
\bibitem{}Wu, K., Chanmugam, G., \& Shaviv, G.  1994, ApJ, 426, 664




\end{thebibliography}
\end{document}